\documentclass[prd,aps,superscriptaddress]{revtex4}

\usepackage{graphicx}
\usepackage{dcolumn}
\usepackage{bm}

\def\k{{\bf k}}

\begin{document}
\voffset = 0.3 true in
\topmargin = -1 true in 

\title{Parity Doubling in the Meson Spectrum}

\author{Eric S. Swanson}
\affiliation{
Department of Physics and Astronomy, University of Pittsburgh,
Pittsburgh PA 15260}
\affiliation{
Jefferson Lab, 12000 Jefferson Ave,
Newport News, VA 23606}

\vskip .5 true cm
\begin{abstract}
A simple argument for the restoration of parity symmetry high in the hadron 
spectrum is presented. The restoration scale is estimated to be 2.5 GeV.
This in turn implies that typical
quark model phenomenology such as scalar confinement or the $^3P_0$ decay
model are only useful for low lying states. Minimal requirements for constructing
more general phenomenologies are discussed. An additional mass degeneracy between 
$J^{++}$ and $ J^{--}$ states 
is shown to occur and an isovector $3^{++}$ state is predicted at roughly 1700 MeV, in 
contradiction with the naive quark model. Similarly, isovector and isoscalar $4^{--}$ states
are predicted at 2000 MeV. Finally, these results imply that Regge 
trajectories must become nonlinear at high spin.

\end{abstract}

\maketitle

\section{Introduction}
Spontaneous chiral symmetry breaking is one of the celebrated phenomena of low energy
QCD. Its immediate consequence is a triplet (or octet) of light mesons whose existence
is of central importance to much of nuclear and hadronic physics. Here I address the 
issue of the restoration of this symmetry  (or rather, the irrelevance of its breaking) 
 high in the
hadron spectrum. A simple argument in support of this notion is presented 
and some implications are discussed. An important conclusion is that the
nonrelativistic quark model must fail at some mass scale, and this scale can be
surprisingly low.

The standard proof that chiral pions are massless 
supposes that the action and measure of QCD are
invariant under a continuous chiral transformation. This implies that the second
derivative of the  effective potential is zero at the classical field minimum which in
turn implies that the connected one-particle-irreducible propagator has a pole at
zero momentum. The number of such poles is equal to the number of independent broken
symmetries. 
Thus only a single triplet
of pions is predicted and Goldstone's theorem has nothing to say about the
excited pion spectrum. Indeed, the relatively close masses of the $\pi'$ (1300 MeV)
and the $\rho'$ (1450 MeV) is a strong indication that the $\pi'$  may be considered
an ordinary $q\bar q$ bound state. 
It is thus tempting to speculate that chiral symmetry is irrelevant high in the
hadron spectrum. Certainly, one expects this to be the case once the typical
momentum transfers in hadronic bound states exceed the chiral symmetry breaking scale,
$\langle p \rangle >> \Lambda_{\chi SB} \approx$ 1 GeV\cite{orsay}, since perturbative
QCD knows nothing about vacuum structure.
This expectation has been recently been substantiated by Cohen and Glozman who  use
quark hadron duality and the  operator product expansion to
argue that chiral symmetry should be restored high in the baryon spectrum\cite{CG}.
And in fact, experimental indications of parity restoration have been reported for many
years in the baryon spectrum\cite{N}.

In the following it will be argued that parity restoration is a natural consequence
of quark interactions which are chirally symmetric and  relativistic. Of course
QCD obeys both of these constraints. Unfortunately, the majority of current quark
models fail to satisfy these criteria and they must therefore fail to accurately
reproduce the highly excited hadron spectrum.

\section{Chiral Symmetry Restoration in the Hadron Spectrum}

As will become clear, examining the restoration of chiral symmetry requires a relativistic
formalism. It is therefore natural to explore the hadron spectrum with the aid of 
interpolating fields which are constructed in the helicity basis (the reader is reminded that the helicity basis is fully relativistic\cite{MS}). Meson interpolating states (I concentrate on
mesons in the following, however similar arguments apply to baryons) may be written in the
form

\begin{equation}
|JM;\lambda \lambda'\rangle \sim \int d\Omega D^{(J)*}_{M,\lambda-\lambda'} b^\dagger_{\k,\lambda} d^\dagger_{\k,\lambda'} | 0\rangle.
\end{equation}
Phase and normalization factors and the radial wavefunction are left unwritten since they
are irrelevant to the following discussion. Note that these
interpolating fields have zero overlap with quantum number exotic mesons and hence are
only useful in the description of `canonical' mesons. Finally, (possible)
chiral and $U_A(1)$ symmetries
impose a definite isospin structure on low energy QCD. This important point will be 
touched upon again in the conclusions. In the following, perfect isospin symmetry will
be assumed and hence all isospin dependence will be neglected.

Parity and charge conjugation relationships are given by
\begin{equation}
P|JM;\lambda \lambda'\rangle = (-)^{J}|JM;-\lambda -\lambda'\rangle
\end{equation}
and
\begin{equation}
C|JM;\lambda \lambda'\rangle = (-)^{J}|JM;\lambda' \lambda\rangle.
\end{equation}
Thus the complete canonical meson spectrum is spanned by the following interpolating states:

\begin{equation}
|J^{(J+1)(J)}\rangle = {1\over \sqrt{2}}\left( |JM;++\rangle - |JM;--\rangle\right)
\end{equation}
\begin{equation}
|J^{(J+1)(J+1)}\rangle = {1\over \sqrt{2}}\left( |JM;+-\rangle - |JM;-+\rangle\right)
\end{equation}
and

\begin{equation}
|J^{(J)(J)}_1\rangle = {1\over \sqrt{2}}\left( |JM;++\rangle + |JM;--\rangle\right)
\end{equation}
\begin{equation}
|J^{(J)(J)}_2\rangle = {1\over \sqrt{2}}\left( |JM;+-\rangle + |JM;-+\rangle\right).
\end{equation}
The left hand side of these relations refer to $J^{PC}$ quantum numbers where $(J) \equiv (-)^J$.

It is useful to define temporal correlation functions as follows ($T$ may be simply 
taken as an arbitrary parameter):
\begin{equation}
C(J^{(J+1)(J)};T) =  \langle JM;++|{\rm e}^{-HT}|JM;++\rangle - \langle JM;++|{\rm e}^{-HT}|JM;--\rangle
\end{equation}

\begin{equation}
C(J^{(J+1)(J+1)};T) =  \langle JM;+-|{\rm e}^{-HT}|JM;+-\rangle - \langle JM;-+|{\rm e}^{-HT}|JM;+-\rangle.
\end{equation}

In the $J^{(J)(J)}$ channel one must diagonalize the correlator matrix given by

\begin{equation}
C_{ij}(T) = \langle J^{(J)(J)}_i | {\rm e}^{-HT} | J^{(J)(J)}_j \rangle
\end{equation}

The Hamiltonian is understood to be that of QCD in a convenient gauge. Note that the
implicit functional integral may have a nontrivial metric and that the region
of integration may be similarly constrained. However, none of these technical issues 
are relevant to the discussion. 

The occurrence of chiral symmetry breaking has several effects on the behaviour of QCD:
(i) a nonzero condensate appears, (ii) Goldstone modes are generated, (iii) a dynamical
quark mass is generated. Of course these phenomena are all  related by the underlying strong
QCD dynamics. Thus, for example,  Goldstone bosons are fluctuations in the order parameter and are 
bound states of quasiparticle excitations. Furthermore, these quasiparticle excitations
are manifested in QCD via a quark field with a dynamical quark mass, $\mu(p)$, which implicitly 
defines the nontrivial broken vacuum:

\begin{equation}
\psi(p;m_q=0) \to \psi(p;m_q=\mu(p)).
\label{njl}
\end{equation}
The classic work of Nambu and Jona-Lasinio\cite{NJL}
gives an explicit example of the quasiparticle (constituent quark) basis embodied in Eq. \ref{njl}. Similarly, the dynamical quark mass has been related to the chiral condensate by 
Politzer\cite{DP} who used the operator product expansion to obtain

\begin{equation}
\mu(p) \sim {\sigma\over p^2}\langle \bar \psi \psi\rangle, \ p \to \infty
\label{pol}
\end{equation}
where $\sigma = 16 \pi \alpha_s(p)$ modulo logarithms. This behaviour
is very general and is seen in all models of chiral symmetry breaking. Furthermore Politzer
showed that the general result

\begin{equation}
{\mu_f(p)\over \mu_{f'}(p)} \to {m_f \over m_{f'}}, {\mbox{\rm as}}\  p \to \infty,
\label{rat}
\end{equation}
holds and that this relationship is true with or without spontaneous
chiral symmetry breaking. 
Here $m_f$ is the bare mass of a quark of flavour $f$.
Setting $f=u$ or $d$ and $f'=b$ in Eq. \ref{rat} then demonstrates that $\mu_{u/d}(p) \to 0$
as the momentum gets large (the chiral limit is assumed throughout). Of course the same
conclusion follows from Eq. \ref{pol}.

The central point is that the dominant manifestation of spontaneous chiral symmetry
breaking high in the hadronic spectrum is in the quark field expansion and that  the
effective mass appearing in the quark field must approach zero as the average momentum
probed by the quark gets large. But this implies that 
helicity flip transitions are suppressed at momenta which are large compared
to the chiral symmetry breaking scale. (They are also suppressed for massive
quarks when the momenta are much larger than the quark masses.)
Thus helicity states become good
eigenstates (mixing in the $J^{(J)(J)}$ channel is suppressed) and one obtains

\begin{equation}
C(J^{(J+1)(J)};T) = C_{11}(J^{(J)(J)};T); \quad \langle p \rangle >> \Lambda_{\chi SB}
\label{pr}
\end{equation}
and
\begin{equation}
C(J^{(J+1)(J+1)};T) = C_{22}(J^{(J)(J)};T);\quad   \langle p \rangle >> \Lambda_{\chi SB}; \quad J > 0.
\label{pcr}
\end{equation}
Finally, the fact that the correlator is an analytic function in $T$ then implies that 
meson masses obey the same relationships. Thus parity symmetry is restored for highly 
excited canonical mesons. Furthermore a $J^{\pm\pm}$ doublet structure is expected.

At first sight this conclusion may appear implausible since Fock sector mixing such
as generated by gluon exchange or meson loops differ depending on quantum
numbers. Nevertheless, if these effects are generated by chirally symmetric interactions
(and they are)
then the argument given above shows that their effects must become identical in accord
with Eqs. \ref{pr},\ref{pcr} at high momentum scales.
Any interaction, potential or non-potential, local or nonlocal, that is
generated by chirally symmetric local interactions (as is the case in massless QCD) must
obey the same general constraints laid out here.

A further symmetry occurs in the nonrelativistic limit or at high spin, namely

\begin{equation}
M(J^{(J+1)(J)}) = M(J^{(J+1)(J+1)}).
\label{cr}
\end{equation}
This is the familiar statement that spin decouples from dynamics
in the heavy quark limit or when the meson wavefunction is suppressed at the
origin.
The last three relationships 
imply that {\it the conventional meson spectrum falls into degenerate quadruplets}
in the high spin/excitation energy limit.

Since nonrelativistic quark models are based on a different momentum regime
than that considered here (namely $p << m_q$) it is perhaps not surprising
that they can not obtain parity symmetry in the excited spectrum. Such models
typically conserve spin and angular momentum separately (with possible perturbative
mixing). Thus states of opposite parity differ by partial wave and belong to different
multiplets. However, in the relativistic case 
it is a specific and nonperturbative combination of J-1 and J+1 waves which allows the degeneracy to occur\footnote{Specifically, 
$|1\rangle_J =  |(J-1)\rangle/\sqrt{3} - \sqrt{2}|(J+1)\rangle/\sqrt{3}$ and $|2\rangle_J = \sqrt{2}|(J-1)\rangle/\sqrt{3} + |(J+1)\rangle/\sqrt{3}$.}.


It is instructive to estimate precisely where in the spectrum parity is restored.
Such an estimate is permitted by the 
assertion that the dominant manifestation of  chiral symmetry breaking high in the
spectrum is through the dynamical quark mass.  In the nearly degenerate limit, mass
differences may be estimated perturbatively as 
$\delta E \sim \langle \lambda\lambda'| \int \bar \psi \Gamma \psi K \bar \psi \Gamma \psi |
\mu \mu' \rangle$ where $\Gamma$ is a Dirac matrix, $K$ is some interaction
kernel, and $\lambda$, $\lambda'$, $\mu$, $\mu'$ are helicity labels. In the case of
helicity flip transition one finds that
quark spinors occur in pairs and hence a one percent deviation in
meson masses implies that $\langle p \rangle \approx 10 m_q$. For heavy quarks such momenta
are experimentally impractical. For light quarks Politzer's result (Eq. \ref{pol}) may be
used to obtain the estimated parity restoration energy

\begin{equation}
E_{restore} \approx -2 (10 \sigma \langle \bar \psi \psi \rangle)^{1/3} \approx 2.5 \ {\rm GeV}.
\end{equation}
It has been noted that the perturbative regime in which chiral symmetry breaking is
irrelevant may be so high in energy that well-defined resonances may not exist\cite{CG}.
It is therefore satisfying that a relatively low mass scale, for which mesons remain
easily identifiable,  emerges from this analysis.

Unfortunately, little is known of the highly excited meson spectrum. The only
isovector states which permit comparison in the excitation spectrum are the scalars
and pseudoscalars. These may be compared by 
computing the mass 
ratio $(0^{++}-0^{-+})/(0^{++}+0^{-+})$. The results are 75\%, 6.0\%, -2.5\%, -1.0\%, and -0.5\%(-5.8\%)  for the ground state through the fourth excited states respectively\footnote{The following masses from the RPP\cite{PDG} have been employed: $\pi(138)$, $a_0(980)$, $\pi(1300)$,  $a_0(1474)$, $\pi(1800)$. I have used the $f_J(1713)$ mass as an estimate of the $a_0(3S)$ mass. 
Note, however, that Ref.\cite{CB2} claim an $f_0$ at 1770 MeV. Further states are obtained
from Ref. \cite{CB2}: $\pi(2070)$, $\pi(2360)$, $a_0(\sim 2025)$/$f_0(2040)$, and the
$f_0(2337)$. These references also report an $f_0$ at 2102 MeV, which provides the last
ratio in brackets above.}. 
It is clear that relative mass differences become quite small once masses
of roughly 2 GeV are reached.

The only other current option for testing parity doubling is in
low lying high angular momentum states. Again, data is sparse but modest progress can be 
made with $J=3$ and $J=4$ states.
The RPP\cite{PDG} lists the states $\rho_3(1690)$ and $\rho_3(1990)$.  Equations \ref{pr}
and \ref{pcr} lead one to conclude (assuming $E \sim 1700$ MeV is `large') that a 
$3^{++}$ state should 
exist at approximately 1700 MeV with a $3^{+-}$ state at roughly 2000 MeV.
Indeed, there has been a recent report of a $3^{+-}$ state at 2032 MeV from
Crystal Barrel\cite{CB}. The close agreement with expectations lends hope that the predicted
mass of the $3^{++}(1700)$ is reasonably accurate. 
Note that nonrelativistic quark models typically predict that the $3^{++}$ state is nearly
degenerate with the $3^{+-}$\cite{GI}, since they are related by a spin flip.
Thus the discovery of a $3^{++}$ meson with a mass of approximately 1700 MeV 
would be a dramatic confirmation
of the importance of chiral dynamics in QCD and an indication of the limited
range of validity of the nonrelativistic approach.  Of course, similar statements apply to higher spin states. For example, recent analyses of Crystal Barrel data\cite{CB2} report
isoscalar states as follows: $4^{++}(2018)$, $4^{++}(2283)$, and $4^{-+}(2328)$. The near degeneracy between the higher $4^{++}$ state and the $4^{-+}$ is in agreement with Eq. \ref{pr} and leads one to
expect an isoscalar $4^{--}$ state at roughly 2000 MeV. Similarly, isovector states are\cite{CB2} $4^{++}(2005)$, $4^{++}(2255)$, and $4^{-+}(2250)$. Again, an isovector $4^{--}$ is
predicted at 2000 MeV.

A final application of the results presented here concerns the Regge trajectories
of hadronic phenomenology. These are traditionally assumed to be linear in the mass
squared of the hadrons: $M^2 = \alpha' J + J_0$.  However, Eqs. \ref{pr} and \ref{pcr}
imply that {\it Regge trajectories must become nonlinear at moderate spin}. For example
the $\rho$ and $b_1$ trajectories, which may start parallel, must merge past a  restoration scale
of roughly $M^2 > 6$ GeV$^2$, or $J > 5$ since they fall in sequences ($J$ odd)$^{--}$ and 
($J$ odd)$^{+-}$.  Experimentally the gap between the trajectories is 0.93 GeV$^2$ ($J=1$), 
1.27 GeV$^2$ ($J=3$), and roughly 0.73 GeV$^2$ ($J=5$). Thus there is an indication that
the two trajectories are indeed merging.  
Note that this example is not a good one because the lowest lying $\rho_J$ states have been
used to define the $\rho$ trajectory. These may be identified with the helicity states 
$J_1^{(J)(J)}$,
which implies that the comparison is better made with the $a_1$ trajectory.  Thus, this 
test is actually of
the stronger quadruplet degeneracy expected at high $J$, rather than the weaker degeneracies of
Eqs. \ref{pr} and \ref{pcr}. Hopefully, meson spectroscopy will advance to the point
that all low lying mesons up to $J=6$ will discovered. Only then can the systematics of
chiral restoration be explored with some degree of certainty.

\section{Conclusions}

I have argued that the dominant feature of spontaneous chiral symmetry breaking high
in the hadronic spectrum is an effective momentum dependent quark mass present
in the spinors of the quark field. This effective mass serves to redefine the vacuum and
provides and an efficient description of the broken phase of QCD.
Furthermore, the
effective mass will approach zero as the average
momentum probed by the quark gets large. In this case, helicity conservation by chirally
symmetric interactions guarantees that two new degeneracies appear in the meson spectrum,
namely parity and parity-charge conjugation doublets emerge. Furthermore, at moderately high total
angular momentum, these doublets merge into a quadruplet structure, even for low lying states.
Finally, the linear Regge trajectories of hadronic phenomenology must
be regarded as approximations which are only valid low in the spectrum. 

Equations \ref{pr} and \ref{pcr} are consistent with the restoration of SU(2)$_V \times $SU(2)$_A$,
and possibly U(1)$_A \times $U(1)$_V$, symmetry. Determining which requires reinstating
isospin indices in the formalism. Doing so reveals, for example, that the $1^{++}$ state becomes
degenerate with the $1^{--}$ due to restored chiral symmetry. 
Note, however, that it has been argued\cite{CG}
that the same condensates drive spontaneous SU(2) and U(1) symmetry breaking so that both
symmetries are expected to be restored concurrently (if one can neglect the anomalous breaking of 
U(1) symmetry high in the spectrum). Thus distinguishing these symmetry restoration
mechanisms is not necessary. In conclusion, the results presented here are in
complete agreement of those of Ref. \cite{CG}. What is new is the explicit connection to the
constituent quark model, the prediction of quadruplet degeneracy, and the predictions of
the previous section.

It is clear that attempts to model both the low lying and the highly 
excited hadron spectrum must be carried out carefully.  The simplest 
requirement is that any such model be relativistic, otherwise the 
transition to the parity symmetric region can never occur. Furthermore, model
quark interactions should be chirally invariant. Many models employ
a scalar confinement potential. We now know that this can only be regarded as
an effective low energy interaction and that it cannot correctly describe the
highly excited spectrum. A similar example is provided by the phenomenological `$^3P_0$'
strong decay model. The model vertex is sometimes written in the relativistic notation
$V = g \int \bar \psi \psi$. Again, such a chirally noninvariant interaction cannot 
yield correct high excitations and must be regarded as an effective interaction for
low lying states only.

The challenge is to construct new models which are chirally invariant, relativistic, and which still reproduce the successes of earlier models in the low lying spectrum.
This need not be an insurmountable problem, for example 
a model for mesons which incorporates these features and explicitly demonstrates 
the effects discussed
here is developed in Ref. \cite{LS}. Another possible resolution is explored in Ref. \cite{BM}.   
How a new strong decay model may be constructed
which obeys these constraints is discussed in Section V.C of Ref. \cite{ss7}.

\begin{acknowledgments}
I am grateful to Leonid Glozman, Bob Jaffe, and John Terning for fruitful discussions 
and to Leonid Glozman for directing
my attention to Refs. \cite{CB2} and the latter three of Refs. \cite{CG}.
This work was supported by the DOE under contracts DE-FG02-00ER41135 and 
DE-AC05-84ER40150.
\end{acknowledgments}

\end{document}